\documentclass[12pt]{article}
\usepackage{amsfonts}
\def\QATOP#1#2{{#1 \atop #2}}
\def\text#1{\mbox{$#1$}}
\newfont{\bbd}{msbm10 scaled\magstep1}
\renewcommand{\theequation}{\thesection.\arabic{equation}}

\begin{document}
\hfill\vbox{\hbox{hep-th/0011145}}

\begin{center}
{\Large	  \bf  The Lax pairs for elliptic $C_{n}$ and $BC_{n}$
Ruijsenaars-Schneider models \\and their spectral curves}
\end{center}

\begin{center}
Kai Chen\footnote{E-mail: kai@phy.nwu.edu.cn},
Bo-yu Hou\footnote{E-mail: byhou@phy.nwu.edu.cn}\\
{\small \it Institute of Modern Physics, Northwest University, Xi'an
710069, China}\\
Wen-li Yang\footnote{E-mail: wlyang@th.physik.uni-bonn.de}\\
{\small{\it  Physikalisches Institut der Universitat Bonn, Nussallee
12, 53115 Bonn, Germany }}
\end{center}

\begin{abstract}
We study the elliptic $C_{n}$ and $BC_{n}$
Ruijsenaars-Schneider models which is elliptic
generalization of system given in \cite{Che00}. The Lax
pairs for these models are constructed by Hamiltonian
reduction technology. We show that the spectral curves can
be parameterized by the involutive integrals of	 motion for these
models. Taking nonrelativistic limit and scaling limit, we
verify that they lead to the systems corresponding to
Calogero-Moser and Toda types.
\newline\\
{\bf \noindent PACS:} 02.20.+b, 11.10.Lm, 03.80.+r\\
{\bf \noindent Keywords:} Lax pair; Ruijsenaars-Schneider
model; Spectral curve; Hamiltonian reduction\\
\end{abstract}

\section{Introduction}

\setcounter{equation}{0} The Ruijsenaars-Schneider(RS) and
Calogero-Moser(CM) models as integrable many-body models
recently have attracted remarkable attention and have been
extensively studied. They describe one-dimensional
$N$-particle system with pairwise interaction. Their
importance lies in various fields ranging from lattice
models in statistics physics\cite{h1,nksr}, to the field
theory and gauge theory\cite {gm,n}. e.g. to the
Seiberg-Witten theory\cite{bm1}\textit{\ et al.} In particular,
the study of the RS model is of great importance since it is
the integrable relativistic generalization of the corresponding
CM model\cite{r1,r2}.

Recently, the Lax pairs for the CM models in various root system have
been constructed by Olshanetsky and Perelomov\cite{op} using
reduction on symmetric space, further given by Inozemtsev in
\cite{in}. Afterwards, D'Hoker and Phong\cite{hp1} succeeded in	 constructing
the Lax pairs with spectral parameter for each of the finite
dimensional Lie algebra,  as well as the introduction of
untwisted and twisted Calogero-Moser systems. Bordner
\textit{et al.}\cite{bcs,bcs2,bcs3} give two types
universal realization  for the Lax pairs associated
to all of the Lie algebra: the root type and the minimal
type, with and without spectral parameters.
Even for  all of the Coxter group, the construction has been
obtained in \cite{bcs1}.  All of them do not apply the reduction method
for under which condition one will confront some obstruction\cite{hm}
but using pure Lie algebra construction. In \cite {hm},
Hurtubise and Markman utilize so called ``structure group'',
which combines semi-simple group and Weyl group, to
construct CM systems associate with Hitchin system, which in
some degree generalizes the result of Refs.
\cite{hp1,bcs,bcs2,bcs3,bcs1}. Furthermore, the quantum
version of the generalization have been developed in
\cite{bms,kps} at least	 for degenerate potentials of
trigonometric after  the works of Olshanetsky and Perelomov\cite{op1}.

So far as for the RS model, only the Lax
pair of the $A_{N-1}$ type RS model was obtained \cite{r1,nksr,bc,kz,s1,s2} and
succeeded in recovering it  by applying
Hamiltonian reduction procedure on two-dimensional current
group\cite{aru}. Although the
commutative operators for RS model based on various type Lie
algebra have been given by Komori and co-workers
\cite{ko1,ko2}, Diejen\cite{di,di1} and Hasegawa \textit{et
al.} \cite{h1,h2}, the Lax integrability (or Lax pair representation)
of the other type RS model is still an open problem\cite{bm1}
except few  degenerate case\cite{kai3,Che00}.

In the work of Refs. \cite{kai3} and \cite{Che00}, we have
succeeded in constructing the Lax pair for $C_{n}$ and $BC_{n}$ RS systems
only with the degenerate case. The r-matrix structure for them have been
derived by Avan \textit{et al.} \cite{Avan}. In this paper, we study
the Lax pair for the most general  $C_{n}$ and $BC_{n}$ RS models---the
elliptic  $C_{n}$ and $BC_{n}$ RS models.
We shall give the explicit forms of Lax pairs
for these systems by Hamiltonian reduction. We calculate the
spectral curves for these systems, which are showed to be
parameterized by a set of involutive integrals of motion. In
particular, taking their non-relativistic limit and scaling
limit, we shall recover the systems of corresponding CM and
Toda types, respectively. The other various
degenerate cases are also be discussed and the connection
between the Lax pair with spectral parameter and the one
without spectral parameter is remarked.

The paper is organized as follows. The basic materials of
the $A_{N-1}$ RS model are reviewed in Section \ref{ars},
where we propose a Lax pair associating with the Hamiltonian
which has a reflection symmetry with respect to the
particles in the origin. This includes construction of Lax
pair for $A_{N-1}$ RS system together with its symmetry
analysis etc al. The main results are showed in Sections
\ref{Ham} and \ref{lax}. In Section \ref {Ham}, we present
the Lax pairs for the elliptic $C_{n}$ and $BC_{n}$ RS
models by reducing from that of $A_{N-1}$ RS model. The
explicit forms for the Lax pairs are given in Section
\ref{lax}. Section \ref{spec} is devoted to deriving the
spectral curves for these systems and their non-relativistic
counterpart, the Calogero-Moser model and scaling limit of
the Toda model. Section \ref{degen} is to show the various
degenerate limits: the trigonometric, hyperbolic and
rational cases. The last section is a brief summary and
discussion.

\section{The $A_{N-1}$-type Ruijsenaars-Schneider Model}

\setcounter{equation}{0} \label{ars}
As a relativistic-invariant
generalization of the $A_{N-1}$-type nonrelativistic
Calogero-Moser model, the $A_{N-1}$-type
Ruijsenaars-Schneider systems are completely integrable. The
system's integrability was first showed by
Ruijsenaars\cite{r1,r2}. The Lax pair for this model has
been constructed in Refs. \cite{r1,nksr,bc,kz,s1,s2}. Recent progress have showed that the
compactification of higher dimension SUSY Yang-Mills theory
and Seiberg-Witten theory can be described by this
model\cite{bm1}. Instanton correction of the prepotential
associated with the $sl_{2}$ RS system has been calculated
in Ref. \cite {ohta}.

\subsection{Model and equations of motion}

Let us briefly give the basics of this model. In terms of
the canonical variables $p_{i}$, $x_{i}(i,j=1,\ldots ,N)$
enjoying the canonical Poisson bracket
\begin{equation}
\{p_{i},p_{j}\}=\{x_{i},x_{j}\}=0,\mbox{$ \ \ \ \ \ \ \ \ \ \ \ \ \ $}
\{x_{i},p_{j}\}=\delta _{ij},  \label{poisson}
\end{equation}
the Hamiltonian of the $A_{N-1}$ RS system reads as
\begin{equation}
\mathcal{H}_{A_{N-1}}=\sum_{i=1}^{N}\left( e^{p_{i}}\,\prod_{k\neq
i}f(x_{i}-x_{k})+e^{-p_{i}}\,\prod_{k\neq
i}g(x_{i}-x_{k})\right),
\label{anhami}
\end{equation}
where
\begin{eqnarray}
f(x) &:&=\frac{\sigma (x-\gamma )}{\sigma (x)},	 \nonumber
\\ g(x) &:&=f(x)|_{\gamma \rightarrow -\gamma },\ \ \ \ \ \
\ \ \ x_{ik}:=x_{i}-x_{k},  \label{Weyl}
\end{eqnarray}
and $\gamma $ denotes the coupling constant. Here, $\sigma
(x)$ is the Weierstrass $\sigma $-function which is an
entire, odd and quasiperiodic function with a fixed pair of
the primitive quasiperiods $2\omega _{1}$ and $ 2\omega
_{2}$. It can be defined as the infinite product
\[
\sigma (x)=x\prod_{w\in \Gamma \setminus \{0\}}\left(
1-\frac{x}{w}\right)\,\exp \left[ \frac{x}{w}+\frac{1}{2}
(\frac{x}{w})^{2}\right],
\]
where $\Gamma =2\omega _{1}\hbox{\bbd Z}+2\omega
_{3}\hbox{\bbd Z}$ is the corresponding period lattice.
Defining a third dependent quasiperiod $ 2\omega
_{2}=-2\omega _{1}-2\omega _{3}$, one has
\[
\sigma (x+2\omega _{k})=-\sigma (x)e^{2\eta _{k}(x+\omega _{k})},
~~\zeta(x+2\omega _{k})=\zeta (x)+2\,\eta _{k},\mbox{$ \ \ \ \
$}k=1,2,3,
\]
where
\[
\zeta (x)=\frac{\sigma ^{\prime }(x)}{\sigma (x)}\,,\qquad \wp (x)=-\zeta
^{\prime }(x)\,,
\]
and $\eta _{k}=\zeta (\omega _{k})$ satisfy $\eta
_{1}\,\omega _{3}-\eta
_{3}\,\omega _{1}=\frac{\pi i}{2}$.

Notice that in Ref. \cite{r1} Ruijsenaars used another
``gauge'' of the momenta such that two are connected by the
following canonical transformation:
\begin{equation}
x_{i}\longrightarrow x_{i},\ \ \ \ p_{i}\longrightarrow
p_{i}+\frac{1}{2}ln\prod_{j\neq
i}^{N}\frac{f(x_{ij})}{g(x_{ij})}.
\end{equation}
The Lax matrix for this model has the form(for the general
elliptic case)
\begin{equation}
L(\lambda )=\sum_{i,j=1}^{N}\frac{\Phi (x_{i}-x_{j}+\gamma
,\lambda )}{\Phi (\gamma ,\lambda )}\mathrm{\exp
}(p_{j})b_{j}E_{ij},  \label{laxA}
\end{equation}
where
\begin{equation}
\Phi (x,\lambda ):=\frac{\sigma (x+\lambda )}{\sigma (x)\sigma (\lambda )},
\mbox{$\ \ \ \ $}\ b_{j}:=\prod_{k\neq j}f(x_{j}-x_{k}),\ \ \ \mbox{$\ \ \ $}
(E_{ij})_{kl}=\delta _{ik}\delta _{jl}.	 \label{phi}
\end{equation}
and $\lambda$ is the spectral parameter. It is shown in Ref. \cite{s2,Avan1,kns} that the Lax operator
satisfies the quadratic fundamental Poisson bracket
\begin{equation}
\{L_{1},L_{2}\}=L_{1}\,L_{2}\,a_{1}-a_{2}\,L_{1}\,L_{2}+L_{2}\,s_{1}
\,L_{1}-L_{1}\,s_{2}\,L_{2},  \label{quad}
\end{equation}
where $L_{1}=L_{A_{N-1}}\otimes Id,L_{2}=Id\otimes
L_{A_{N-1}}$ and the four matrices read as
\begin{eqnarray}
a_{1} &=&a+w,\quad s_{1}=s-w,  \nonumber \\ a_{2}
&=&a+s-s^{\ast }-w,\quad s_{2}=s^{\ast }+w.
\end{eqnarray}
The forms of $a,s,w$ are
\begin{eqnarray}
a(\lambda ,\mu ) &=&-\zeta (\lambda -\mu
)\sum_{k=1}^{N}E_{kk}\otimes E_{kk}-\sum_{k\neq j}\Phi
(x_{j}-x_{k},\lambda -\mu )E_{jk}\otimes E_{kj},
\nonumber \\
s(\lambda ) &=&\zeta (\lambda )\sum_{k=1}^{N}E_{kk}\otimes
E_{kk}+\sum_{k\neq j}\Phi (x_{j}-x_{k},\lambda
)E_{jk}\otimes E_{kk}, \\ w &=&\sum_{k\neq j}\zeta
(x_{k}-x_{j})E_{kk}\otimes E_{jj}.  \nonumber
\end{eqnarray}
The asterisk symbol means
\begin{equation}
r^{\ast }=\Pi r\Pi \;\;\mbox{$with$}\;\;\Pi
=\sum_{k,j=1}^{N}E_{kj}\otimes E_{jk}.
\end{equation}

Noticing that
\begin{eqnarray}
L(\lambda )^{-1}{}_{ij} &=&\frac{\sigma (\gamma +\lambda
)\,\sigma (\lambda +(N-1)\gamma )}{\sigma (\lambda )\sigma
(\lambda +N\gamma )}  \nonumber \\ &&\times
\sum_{i,j=1}^{N}\frac{\;\Phi (x_{i}-x_{j}-\gamma ,\lambda
+N\gamma ) }{\Phi (-\gamma ,\lambda +N\gamma )}\mathrm{\exp
}(-p_{i})b_{j}^{^{\prime }}E_{ij},  \label{laxinverse} \\
\mbox{$\ \ $}b_{j}^{^{\prime }} &:&=\prod_{k\neq j}g(x_{j}-x_{k}),
\end{eqnarray}
(the proof of the above identity is sketched in the
appendix) one can get the characteristic polynomials of $L_{A_{N-1}}$
(Ref. \cite{r3})
\begin{equation}
\det (L(\lambda )-v\cdot Id)=\sum_{j=0}^{N}\Phi (\gamma ,\lambda
)^{-j}(-v)^{N-j}\frac{\mathcal{H}_{j}^{+}}{\sigma
^{j}(\gamma )}\;\times \;
\frac{\sigma (\lambda +j\gamma )}{\sigma (\lambda )},  \label{poly}
\end{equation}
and that of $L_{A_{N-1}}^{-1}$ by using formula given in Eq.
(\ref{Inll})
\begin{eqnarray}
&&\det (\frac{\sigma (\lambda )\sigma (\lambda -N\gamma
)}{\sigma (\lambda
-\gamma )\,\sigma (\lambda -(N-1)\gamma )}\times L(\lambda -N\gamma
)^{-1}-v\cdot Id)  \nonumber \\ &=&\sum_{j=0}^{N}\Phi
(-\gamma ,\lambda )^{-j}(-v)^{N-j}\times \frac{(
\mathcal{H}_{j}^{-})}{\sigma ^{j}(-\gamma )}\;\frac{\sigma (\lambda -j\gamma
)}{\sigma (\lambda )},	\label{Ipoly}
\end{eqnarray}
where$(\mathcal{H}_{0}^{\pm
})_{A_{N-1}}=(\mathcal{H}_{N}^{\pm })_{A_{N-1}}=1 $ and
\begin{eqnarray}
(\mathcal{H}_{i}^{+})_{A_{N-1}}
&=&\sum_{\QATOP{J\subset \{1,\ldots ,N\} }{\left| J\right|
=i}}\exp \left(
\sum_{j\in J}p_{j}\right) \,\prod_{\QATOP{j\in J }{k\in \{1,\ldots
,N\}\setminus J}}f(x_{j}-x_{k}), \\
(\mathcal{H}_{i}^{-})_{A_{N-1}}
&=&\sum_{\QATOP{J\subset \{1,\ldots ,N\} }{\left| J\right|
=i}}\exp \left(
\sum_{j\in J}-p_{j}\right) \,\prod_{\QATOP{j\in J }{k\in \{1,\ldots
,N\}\setminus J}}g(x_{j}-x_{k}).
\end{eqnarray}

Define
\begin{equation}
(\mathcal{H}_{i})_{A_{N-1}}=(\mathcal{H}_{i}^{+})_{A_{N-1}}+(\mathcal{H}
_{i}^{-})_{A_{N-1}},  \label{hamiset}
\end{equation}
from the fundamental Poisson bracket Eq. (\ref{quad}), we can
verify that
\begin{equation}
\{(\mathcal{H}_{i})_{A_{N-1}},(\mathcal{H}_{j})_{A_{N-1}}\}=\{(\mathcal{H}
_{i}^{\varepsilon })_{A_{N-1}},(\mathcal{H}_{j}^{\varepsilon ^{^{\prime
}}})_{A_{N-1}}\}=0,\ \ \ \ \ \ \varepsilon ,\varepsilon
^{^{\prime }}=\pm ,\
\ \ \ i,j=1,\ldots ,N.	\label{aninv}
\end{equation}
In particular, the Hamiltonian Eq. (\ref{anhami}) can be
rewritten as

\begin{eqnarray}
\mathcal{H}_{A_{N-1}} &\equiv &\mathcal{H}_{1}=(\mathcal{H}
_{1}^{+})_{A_{N-1}}+(\mathcal{H}_{1}^{-})_{A_{N-1}}=
\sum_{j=1}^{N}(e^{p_{j}}b_{j}+e^{-p_{j}}b_{j}^{^{\prime }})  \nonumber \\
&=&Tr(L(\lambda )+\frac{\sigma (\lambda )\sigma (\lambda
+N\gamma )}{\sigma (\gamma +\lambda )\,\sigma (\lambda
+(N-1)\gamma )}L(\lambda )^{-1})
\end{eqnarray}
It should be remarked the set of integrals of motion Eq.
(\ref{hamiset}) have a reflection symmetry which is the key
property for the later reduction to $C_{n}$ and $BC_{n}$
cases. i.e. if we set

\begin{equation}
p_{i}\longleftrightarrow -p_{i},\mbox{$ \ \ \ \ \
$}x_{i}\longleftrightarrow
-x_{i},	 \label{sym}
\end{equation}
then the Hamiltonians flows $(\mathcal{H}_{i})_{A_{N-1}}$
are invariant with respect to this symmetry.

The canonical equations of motion associated with the
Hamiltonian flows $
\mathcal{H}_{1}^{+}$  in its generic (elliptic) form
read
\begin{equation}
\ddot{x}_{i}=\sum_{j\neq i}\dot{x}_{i}\dot{x}_{j}(V(x_{ij})-V(x_{ji})),\ \
i=1,\ldots ,N\ ,
\end{equation}
where the potential $V(x)$ is given by
\begin{equation}
V(x)=\zeta (x)-\zeta (x+\lambda ),
\end{equation}
in which $\zeta (x)=\frac{\sigma ^{\prime }(x)}{\sigma
(x)}$. Here, $x_{i}=x_{i}(t)$ ,
$p_{i}=p_{i}(t)$ and the superimposed dot denotes
$t$-differentiation.

\subsection{The construction of Lax pair for the $A_{N-1}$ RS model}

As for the $A_{N-1}$ RS model, a generalized Lax pair has
been given in Refs. \cite{nksr,r1} and \cite{bc}-\cite{s2}. But
there is a common character that the time-evolution of the
Lax matrix $L_{A_{N-1}}$ is associated with the Hamiltonian
$(\mathcal{H}_{1}^{+})_{A_{N-1}}$. We will see in section
\ref{Ham} that the Lax pair cannot reduce from that kind of
forms directly. Instead, we give a new Lax pair in which the
evolution of $L_{A_{N-1}}$ is associated with the
Hamiltonian $\mathcal{H}_{A_{N-1}}$,
\begin{equation}
\dot{L}_{A_{N-1}}=\{L_{A_{N-1}},\mathcal{H}_{A_{N-1}}\}=\lbrack
M_{A_{N-1}},L_{A_{N-1}}\rbrack ,  \label{laxeq}
\end{equation}
where $M_{A_{N-1}}$ can be constructed with the help of
$(r,s)$ matrices as follows
\begin{equation}
M_{_{A_{N-1}}}=Tr_{2}((s_{1}-a_{2})(1\otimes (L(\lambda
)-\frac{\sigma (\lambda )\sigma (\lambda +N\gamma )}{\sigma
(\gamma +\lambda )\,\sigma (\lambda +(N-1)\gamma )}L(\lambda
)^{-1}))),
\end{equation}
The explicit expression of entries for $M_{A_{N-1}}$ is
\begin{eqnarray}
M_{ij} &=&\Phi (x_{ij},\lambda )e^{p_{j}}b_{j}-\Phi
(x_{ij},\lambda +N\gamma )e^{-p_{i}}b_{j}^{^{\prime }},~~i\neq j, \\
M_{ii} &=&\Big(\zeta (\lambda )+\zeta (\gamma
)\Big)e^{p_{i}}b_{i}-\Big(
\zeta (\lambda +\gamma )-\zeta (\gamma )\Big)e^{-p_{i}}b_{i}^{^{
\prime }} \\
&&+\sum_{j\neq i}\Big( \Big(\zeta (x_{ij}+\gamma )-\zeta
(x_{ij})\Big) e^{p_{j}}b_{j}  \nonumber \\
&&+\frac{\Phi
(x_{ji}+\gamma ,\lambda )}{\Phi (\gamma ,\lambda )}\Phi
(x_{ij},\lambda +N\gamma )e^{-p_{i}}b_{j}^{^{\prime
}}\Big).
\end{eqnarray}

\section{ Hamiltonian Reduction of $C_{n}$ and $BC_{n}$ RS
Models from $A_{N-1}$-type Ones}
\setcounter{equation}{0} \label{Ham}
Let us first mention some
results about the integrability of Hamiltonian
(\ref{anhami}). In Ref. \cite {r2} Ruijsenaars demonstrated
that the symplectic structure of the $C_{n}$- and
$BC_{n}$-type RS systems can be proved integrable by
embedding their phase space to a submanifold of the
$A_{2n-1}$ and $A_{2n}$ type RS ones, respectively, while in
Refs. \cite{di,di1} and \cite{ko2}, Diejen and Komori,
respectively, gave a series of commuting difference
operators which led to their quantum integrability. However,
there are not any results about their Lax representations so
far except for
the special degenerate	case \cite{kai3,Che00}. In this section, we
concentrate our treatment on the exhibition of the explicit
forms for general $C_{n}$ and $BC_{n}$ RS systems.

For the convenience of analysis of symmetry, let us first
give vector representation of $A_{N-1}$ Lie algebra.
Introducing an $N$ dimensional orthonormal basis of
${\mathbb R}^{N}$,
\begin{equation}
e_{j}\cdot e_{k}=\delta _{j,k},\quad j,k=1,\ldots ,N.
\end{equation}
Then the sets of roots and vector weights of $A_{N-1}$ type  are:
\begin{eqnarray}
\Delta &=&\{e_{j}-e_{k}:\quad j,k=1,\ldots ,N\},\quad  \label{anroot} \\
\Lambda &=&\{e_{j}:\quad j=1,\ldots ,N\}.  \label{anwei}
\end{eqnarray}

The dynamical variables are canonical coordinates
$\{x_{j}\}$ and their canonical conjugate momenta
$\{p_{j}\}$ with the Poisson brackets of Eq. (
\ref{poisson}) . In a general sense, we denote them by $N$
dimensional vectors $x$ and $p$,
\[
x=(x_{1},\ldots ,x_{N})\in {\mathbb R}^{N},\quad
p=(p_{1},\ldots ,p_{N})\in {
\mathbb R}^{N},\quad
\]
so that the scalar products of $x$ and $p$ with the roots
$\alpha \cdot x$, $ p\cdot \beta $, etc. can be defined.
The Hamiltonian of Eq. (\ref{anhami}) \ can be rewritten as
\begin{equation}
\mathcal{H}_{A_{N-1}}=\sum_{\mu \in \Lambda }\left( \exp \left(
\mu \cdot p\right) \,\prod_{\Delta \ni \beta =\mu -\nu }f(\beta \cdot
x){+}\exp \left(
-\mu \cdot p\right) \,\prod_{\Delta \ni \beta =-\mu +\nu }g(\beta
\cdot x)\right) ,
\end{equation}
Here, the condition $\Delta \ni \beta =\mu -\nu $ means that
the summation is over roots $\beta $ such that for $\exists
\nu \in \Lambda $

\begin{equation}
\mu -\nu =\beta \in \Delta .
\end{equation}
So does for $\Delta \ni \beta =-\mu +\nu .$

\subsection{The $C_{n}$ model}

The set of $C_{n}$ roots consists of two parts, long roots
and short roots:
\begin{equation}
\Delta _{C_{n}}=\Delta _{L}\cup \Delta _{S},  \label{cnroot}
\end{equation}
in which the roots are conveniently expressed in terms of an
orthonormal basis of ${\mathbb R}^{n}$:
\begin{eqnarray}
\Delta _{L} &=&\{\pm 2e_{j}:\quad \qquad \ \ j=1,\ldots ,n\},  \nonumber \\
\Delta _{S} &=&\{\pm e_{j}\pm e_{k},:\quad j,k=1,\ldots ,n\}.
\end{eqnarray}
In the vector representation, vector weights $\Lambda $ are

\begin{equation}
\Lambda _{C_{n}}=\{e_{j},-e_{j}:\quad j=1,\ldots ,n\}.
\end{equation}
The Hamiltonian of the $C_{n}$ model is given by
\begin{equation}
\mathcal{H}_{C_{n}}=\frac{1}{2}\sum_{\mu \in \Lambda _{C_{n}}}
\left( \exp \left( \mu \cdot p\right) \,\prod_{\Delta _{C_{n}}
\ni \beta =\mu -\nu}f(\beta \cdot x){+}\exp \left( -\mu \cdot p\right)
\,\prod_{\Delta
_{C_{n}}\ni \beta =-\mu +\nu }g(\beta \cdot x)\right) .	 \label{cnhami}
\end{equation}
From the above-mentioned data, we notice that either for
$A_{N-1}$ or $C_{n}$ Lie algebra, any root $\alpha \in
\Delta $ can be constructed in terms with vector weights as
$\alpha =\mu -\nu $ where $\mu ,\nu \in \Lambda .$ By simple
comparison of representation between $A_{N-1}$ or $C_{n}$,
one can find that if replacing $e_{j+n}$ with $-e_{j}$ in
the vector weights of $A_{2n-1}$ algebra, we can obtain
the vector weights of $C_{n}$ one. This also holds for the
corresponding roots. This gives us a hint that it is
possible to get the $C_{n}$ model by this kind of reduction.

For the $A_{2n-1}$ model let us set restrictions on the vector
weights with
\begin{equation}
e_{j+n}+e_{j}=0,\mbox{\ \ \  for \ \ \ }j=1,\ldots ,n,
\end{equation}
which correspond to the following constraints on the phase
space of the $A_{2n-1}$-type RS model with
\begin{eqnarray}
G_{i} &\equiv &(e_{i+n}+e_{i})\cdot x=x_{i}+x_{i+n}=0,
\nonumber \\ G_{i+n} &\equiv &(e_{i+n}+e_{i})\cdot
p=p_{i}+p_{i+n}=0,\ \ i=1,\ldots ,n,
\label{cncon}
\end{eqnarray}

\noindent Following Dirac's method\cite{Dirac}, we can show
\begin{equation}
\{G_{i},\mathcal{H}_{A_{2n-1}}\}\simeq 0,\mbox{\ \ \ for \ \ \ \ }\forall
i\in \{1,\ldots ,2n\},	\label{cnfirst}
\end{equation}
i.e. $\mathcal{H}_{A_{2n-1}}$ is the first class Hamiltonian
corresponding to the constraints in Eq. (\ref{cncon}). Here
the ``weak equal" symbol $\simeq $ represents that, only after calculating
the result of left-hand side of the identity, could we use
the conditions of constraints. It should be pointed out that
the most necessary condition ensuring Eq. (\ref{cnfirst}) is
the symmetry property of formula (\ref{sym}) for the
Hamiltonian Eq. (\ref {anhami}). So for an arbitrary
dynamical variable $A$, we have

\begin{eqnarray}
\dot{A} &=&\{A,\mathcal{H}_{A_{2n-1}}\}_{D}=\{A,\mathcal{H}
_{A_{2n-1}}\}-\{A,G_{i}\}\Delta _{ij}^{-1}\{G_{j},\mathcal{H}_{A_{2n-1}}\}
\nonumber \\
&\simeq &\{A,\mathcal{H}_{A_{2n-1}}\},\qquad \ \ \ \ \ \ \ \
\ \ \ \ \ \ \ \
\ \ \ \ \ \ \ \ \ \ \ \ \ \ \ \ \ \ \ i,j=1,\ldots ,2n,	 \label{cnevo}
\end{eqnarray}
where
\begin{equation}
\Delta _{ij}=\{G_{i},G_{j}\}=2\left(
\begin{array}{cc}
0 & Id \\
-Id & 0
\end{array}
\right) ,
\end{equation}
and $\{,\}_{D}$ denote the Dirac bracket. By straightforward
calculation, we have the nonzero Dirac brackets of
\begin{eqnarray}
\{x_{i},p_{j}\}_{D} &=&\{x_{i+n},p_{j+n}\}_{D}=\frac{1}{2}\delta _{i,j},
\nonumber \\
\{x_{i},p_{j+n}\}_{D} &=&\{x_{i+n},p_{j}\}_{D}=-\frac{1}{2}\delta _{i,j}.
\label{DBra}
\end{eqnarray}
Using the above-mentioned data together with the fact that
$\mathcal{H}
_{A_{N-1}}$ is the first class Hamiltonian[see Eq. (\ref{cnfirst})], we can
directly obtain a Lax representation of the $C_{n}$ RS model
by imposing constraints $G_{k}$ on Eq. (\ref{laxeq})
\begin{eqnarray}
\{L_{A_{2n-1}},\mathcal{H}_{A_{2n-1}}\}_{D} &=&\{L_{A_{2n-1}},\mathcal{H}
_{A_{2n-1}}\}|_{G_{k},k=1,...,2n},  \nonumber \\
&=&\lbrack M_{A_{2n-1}},L_{A_{2n-1}}\rbrack
|_{G_{k},k=1,...,2n}=\lbrack M_{C_{n}},L_{C_{n}}\rbrack , \\
\{L_{A_{2n-1}},\mathcal{H}_{A_{2n-1}}\}_{D} &=&\{L_{C_{n}},\mathcal{H}
_{C_{n}}\},  \label{dicn}
\end{eqnarray}
where
\begin{eqnarray}
\mathcal{H}_{C_{n}} &=&\frac{1}{2}\mathcal{H}_{A_{2n-1}}|_{G_{k},k=1,...,2n},
\nonumber \\
L_{C_{n}} &=&L_{A_{2n-1}}|_{G_{k},k=1,...,2n},
\label{reduce} \\ M_{C_{n}}
&=&M_{A_{2n-1}}|_{G_{k},k=1,...,2n},  \nonumber
\end{eqnarray}
so that
\begin{equation}
\dot{L}_{C_{n}}=\{L_{C_{n}},\mathcal{H}_{C_{n}}\}=\lbrack
M_{C_{n}},L_{C_{n}}\rbrack .  \label{cnLax}
\end{equation}

Nevertheless, the $(\mathcal{H}_{1}^{+})_{A_{N-1}}$ is not
the first class Hamiltonian, so the Lax pair given by many
authors previously cannot reduce to the $C_{n}$ case
directly in this way.

\subsection{The $BC_{n}$ model}

The $BC_{n}$ root system consists of three parts: long,
middle and short roots:
\begin{equation}
\Delta _{BC_{n}}=\Delta _{L}\cup \Delta \cup \Delta _{S},  \label{bcnroot}
\end{equation}
in which the roots are conveniently expressed in terms of an
orthonormal basis of ${\mathbb R}^{n}$:
\begin{eqnarray}
\Delta _{L} &=&\{\pm 2e_{j}:\qquad \quad \ \ j=1,\ldots ,n\},  \nonumber \\
\Delta &=&\{\pm e_{j}\pm e_{k}:\quad j,k=1,\ldots ,n\}, \\
\Delta _{S} &=&\{\pm e_{j}:\qquad \quad \ \ \ \ j=1,\ldots ,n\}.  \nonumber
\end{eqnarray}
In the vector representation, vector weights $\Lambda $ can
be

\begin{equation}
\Lambda _{BC_{n}}=\{e_{j},-e_{j},0:\quad j=1,\ldots ,n\}.
\end{equation}
The Hamiltonian of $BC_{n}$ model is given by
\begin{equation}
\mathcal{H}_{BC_{n}}=\frac{1}{2}\sum_{\mu \in \Lambda _{BC_{n}}}\left( \exp
\left( \mu \cdot p\right) \,\prod_{\Delta _{BC_{n}}\ni \beta =\mu -\nu
}f(\beta \cdot x){+}\exp \left( -\mu \cdot p\right)
\,\prod_{\Delta
_{BC_{n}}\ni \beta =-\mu +\nu }g(\beta \cdot x)\right) .  \label{bcnhami}
\end{equation}
By similar comparison of representations between $A_{N-1}$
or $BC_{n}$, one can find that if replacing $e_{j+n}$ with
$-e_{j}$ and $e_{2n+1}$ with $0$ in the vector weights of
the $A_{2n}$ Lie algebra, we can obtain the vector weights
of the $BC_{n}$ one. The same holds for the corresponding
roots. So by the same procedure as $C_{n}$ model, we could
get the Lax representation of the $BC_{n}$ model.

For the $A_{2n}$ model, we set restrictions on the vector
weights with

\begin{eqnarray}
e_{j+n}+e_{j} &=&0,\mbox{\ \ \ for \ \ \ }j=1,\ldots ,n,
\nonumber \\ e_{2n+1} &=&0,
\end{eqnarray}
which correspond to the following constraints on the phase
space of $A_{2n}$
-type RS model with

\begin{eqnarray}
G_{i}^{^{\prime }} &\equiv &(e_{i+n}+e_{i})\cdot
x=x_{i}+x_{i+n}=0,
\nonumber \\
G_{i+n}^{^{\prime }} &\equiv &(e_{i+n}+e_{i})\cdot
p=p_{i}+p_{i+n}=0,\ \ i=1,\ldots ,n,  \nonumber \\
G_{2n+1}^{^{\prime }} &\equiv &e_{2n+1}\cdot x=x_{2n+1}=0,
\nonumber \\ G_{2n+2}^{^{\prime }} &\equiv &e_{2n+1}\cdot
p=p_{2n+1}=0.  \label{bcncon}
\end{eqnarray}

\noindent Similarly, we can show

\begin{equation}
\{G_{i},\mathcal{H}_{A_{2n}}\}\simeq 0,\mbox{\ \ \ for \ \ \ \ }\forall i\in
\{1,\ldots ,2n+1,2n+2\}.
\end{equation}
i.e. $\mathcal{H}_{A_{2n}}$ is the first class Hamiltonian
corresponding to the above-mentioned constraints Eq.
(\ref{bcncon}). So $L_{BC_{n}}$ and $M_{BC_{n}}$ can be
constructed as follows

\begin{eqnarray}
L_{BC_{n}} &=&L_{A_{2n}}|_{G_{k}^{^{\prime }},k=1,...,2n+2},
\nonumber \\ M_{BC_{n}} &=&M_{A_{2n}}|_{G_{k}^{^{\prime
}},k=1,...,2n+2},
\end{eqnarray}
while \ $\mathcal{H}_{BC_{n}}$ is

\begin{equation}
\mathcal{H}_{BC_{n}}=\frac{1}{2}\mathcal{H}_{A_{2n}}|_{G_{k},k=1,...,2n+2},
\end{equation}
due to the similar derivation of Eqs.
(\ref{cnevo}-\ref{cnLax}).

\section{Lax Representations of the $C_{n}$ and $BC_{n}$ RS Models}

\label{lax} \setcounter{equation}{0}

\subsection{The $C_{n}$ model}

The Hamiltonian of the $C_{n}$ RS system is Eq.
(\ref{cnhami}), so the canonical equations of motion are

\begin{eqnarray}
\dot{x_{i}} &=&\{x_{i},\mathcal{H}\}=e^{p_{i}}b_{i}-e^{-p_{i}}b_{i}^{^{
\prime }},  \label{mocn1} \\
\dot{p_{i}} &=&\{p_{i},\mathcal{H}\}=\sum_{j\neq i}^{n}\Big(e^{p_{j}}b_{j}
\big(h(x_{ji})-h(x_{j}+x_{i})\big)  \nonumber \\
&&+e^{-p_{j}}b_{j}^{^{\prime
}}\big(\hat{h}(x_{ji})-\hat{h}(x_{j}+x_{i})\big)
\Big)  \nonumber \\
&&-e^{p_{i}}b_{i}\Big(2h(2x_{i})+\sum_{j\neq i}^{n}\big(
h(x_{ij})+h(x_{i}+x_{j})\big)\Big)  \nonumber \\
&&-e^{-p_{i}}b_{i}^{^{\prime
}}\Big(2\hat{h}(2x_{i})+\sum_{j\neq i}^{n}\big(
\hat{h}(x_{ij})+\hat{h}(x_{i}+x_{j})\big)\Big),	 \label{mocn2}
\end{eqnarray}
where
\begin{eqnarray}
h(x):= &&\frac{d\ln f(x)}{dx},\mbox{$ \ \ \ \ \ \ \ \ \ \ \
$}\hat{h}(x):=
\frac{d\ln g(x)}{dx},  \nonumber \\
b_{i} &=&f(2x_{i})\,\prod_{k\neq
i}^{n}\Big(f(x_{i}-x_{k})f(x_{i}+x_{k})\Big), \\
b_{i}^{^{\prime }} &=&g(2x_{i})\,\prod_{k\neq i}^{n}\Big(
g(x_{i}-x_{k})g(x_{i}+x_{k})\Big).  \nonumber
\end{eqnarray}

The Lax matrix for the $C_{n}$ RS model can be written in
the following form
\begin{equation}
(L_{C_{n}})_{\mu \nu }=e^{\nu \cdot p}b_{\nu }\frac{\Phi
((\mu -\nu )\cdot x+\gamma ,\lambda )}{\Phi (\gamma ,\lambda
)},  \label{Lmatrix}
\end{equation}
which is a $2n\times 2n$ matrix whose indices are labelled
by the vector weights, denoted by \ $\mu ,\nu \in \Lambda
_{C_{n}}$, $M_{C_{n}}$ can be written as

\begin{equation}
M_{C_{n}}=D+Y,	\label{Mmatrix}
\end{equation}
where $D$ denotes the diagonal part and $Y$ denotes the
off-diagonal part
\begin{eqnarray}
Y_{\mu \nu } &=&e^{\nu \cdot p}b_{\nu }\Phi (x_{\mu \nu
},\lambda )
\nonumber \\
&&+e^{-\mu \cdot p}b_{\nu }^{^{\prime }}\Phi (x_{\mu \nu
},\lambda +N
\mbox{$
$}\gamma ),  \label{elliY} \\ D_{\mu \mu } &=&\Big(\zeta
(\lambda )+\zeta (\gamma )\Big)e^{\mu \cdot p}b_{\mu
}-\Big(\zeta (\lambda +\gamma )-\zeta (\gamma )\Big)e^{-\mu
\cdot p}b_{\mu }^{^{\prime }}  \nonumber \\ &&+\sum_{\nu
\neq \mu }\Big(\Big(\zeta (x_{\mu \nu }+\gamma )-\zeta
(x_{\mu
\nu })\Big)e^{\nu \cdot p}b_{\nu }  \nonumber \\
&&+\frac{\Phi (x_{\nu \mu }+\gamma ,\lambda )}{\Phi (\gamma
,\lambda )}\Phi (x_{\mu \nu },\lambda +N\gamma )e^{-\mu
\cdot p}b_{\nu }^{^{\prime }}\Big)
\label{elliD}
\end{eqnarray}
and
\begin{eqnarray}
b_{\mu } &=&\prod_{\Delta _{C_{n}}\ni \beta =\mu -\nu
}f(\beta \cdot x),
\nonumber \\
b_{\mu }^{^{\prime }} &=&\prod_{\Delta _{C_{n}}\ni \beta
=\mu -\nu }g(\beta
\cdot x),  \label{cnaux} \\
x_{\mu \nu } &:&=(\mu -\nu )\cdot x.  \nonumber
\end{eqnarray}

The $L_{C_{n}},M_{C_{n}}$ satisfies the Lax equation

\begin{equation}
\dot{L}_{C_{n}}=\{L_{C_{n}},\mathcal{H}_{C_{n}}\}=\lbrack
M_{C_{n}},L_{C_{n}}\rbrack ,
\end{equation}
which equivalent to the equations of motion Eq.
(\ref{mocn1}) and Eq. (\ref {mocn2}). The Hamiltonian
$\mathcal{H}_{C_{n}}$ can be rewritten as the trace of
$L_{C_{n}}$

\begin{equation}
\mathcal{H}_{C_{n}}=trL_{C_{n}}=\frac{1}{2}\sum_{\mu \in \Lambda
_{C_{n}}}(e^{\mu \cdot p}b_{\mu }+e^{-\mu \cdot p}b_{\mu }^{^{\prime }}).
\end{equation}

\subsection{The $BC_{n}$ model}

The Hamiltonian of the $BC_{n}$ model is expressed in Eq.
(\ref{bcnhami}), so the canonical equations of motion are
\begin{eqnarray}
\dot{x_{i}} &=&\{x_{i},\mathcal{H}\}=e^{p_{i}}b_{i}-e^{-p_{i}}b_{i}^{^{
\prime }},  \label{mobcn1} \\
\dot{p_{i}} &=&\{p_{i},\mathcal{H}\}=\sum_{j\neq i}^{n}\Big(e^{p_{j}}b_{j}
\big(h(x_{ji})-h(x_{j}+x_{i})\big)  \nonumber \\
&&+e^{-p_{j}}b_{j}^{^{\prime
}}\big(\hat{h}(x_{ji})-\hat{h}(x_{j}+x_{i})\big)
\Big)  \nonumber \\
&&-e^{p_{i}}b_{i}\Big(h(x_{i})+2h(2x_{i})+\sum_{j\neq
i}^{n}\big(h(x_{ij})+h(x_{i}+x_{j})\big)\Big)  \nonumber
\\ &&-e^{-p_{i}}b_{i}^{^{\prime
}}\Big(\hat{h}(x_{i})+2\hat{h} (2x_{i})+\sum_{j\neq
i}^{n}\big(\hat{h}(x_{ij})+\hat{h}(x_{i}+x_{j})\big)
\Big)  \nonumber \\
&&-b_{0}\Big(h(x_{i})+\hat{h}(x_{i})\Big),  \label{mobcn2}
\end{eqnarray}
where

\begin{eqnarray}
b_{i} &=&f(x_{i})f(2x_{i})\,\prod_{k\neq i}^{n}\Big(
f(x_{i}-x_{k})f(x_{i}+x_{k})\Big),  \nonumber \\
b_{i}^{^{\prime }} &=&g(x_{i})g(2x_{i})\,\prod_{k\neq
i}^{n}\Big( g(x_{i}-x_{k})g(x_{i}+x_{k})\Big), \\ b_{0}
&=&\prod_{i=1}^{n}f(x_{i})g(x_{i}).  \nonumber
\end{eqnarray}

The Lax pair for the $BC_{n}$ RS model can be constructed as
the form of Eq. (\ref{Lmatrix})-(\ref{cnaux}) where one
should replace the matrices labels with $\mu ,\nu \in
\Lambda _{BC_{n}},$ and roots with $\beta \in \Delta
_{BC_{n}}.$

The Hamiltonian $\mathcal{H}_{BC_{n}}$ can also be rewritten
as the trace of $L_{BC_{n}}$

\begin{equation}
\mathcal{H}_{BC_{n}}=trL_{BC_{n}}=\frac{1}{2}\sum_{\mu \in \Lambda
_{BC_{n}}}(e^{\mu \cdot p}b_{\mu }+e^{-\mu \cdot p}b_{\mu }^{^{\prime }}).
\end{equation}

\section{Spectral Curves of the $C_{n}$ and $BC_{n}$ RS Systems}

\setcounter{equation}{0} \label{spec} It is recently pointed out in \cite
{n,bm1,mir,mir1} that the $SU(N)$ RS model is related to five
dimensional gauge theories. In the context of Seiberg-Witten
theory, the elliptic RS integrable system can be linked with
the relevant low energy effective action when a
compactification from five dimension to four dimension is
imposed with all of the fields belong to the adjoint
representation of $SU(N) $ gauge group\cite{bm1}. More
evidence for this correspondence between SYM and RS model is
depicted by calculating instanton correction of prepotential
for $SU(2) $ Seiberg-Witten theory in \cite{ohta}.

As for the spectral curve and its relation with the
Seiberg-Witten spectral curve, much progress have been made
in the correspondence of ``Calogero-Moser integrable
theories and gauge theories ''. See recent review in
\cite{hp2}, \cite{Marbook} and references therein. In this
section we will give the spectral curves for $C_{n}$ and
$BC_{n}$ systems, which are shown to be parameterized by the
integrals of motion of the corresponding system. We will also
see that the elliptic Calogero-Moser, Toda (affine and
non-affine) ones are particular limits of these systems.

\subsection{Spectral curve of the $C_{n}$ RS system}

Given the Lax operator with spectral parameter for the
Calogero-Moser system and of the RS system associated with
Lie algebras $\mathcal{G}$, the spectral curve for the given
system is defined as

\begin{equation}
\Gamma \ :\ R(v,\lambda )=~\det (L(\lambda )-v\cdot Id)\equiv 0.
\end{equation}
It is naturally that the function $R(v,z)$ is invariant
under time evolution,
\begin{equation}
{\frac{d}{dt}}R(v,\lambda )=\{\mathcal{H},R(v,\lambda )\}=0.
\end{equation}
Thus, $R(v,\lambda )$ must be a function of only the $n$
independent integrals of motion, which in super-Yang-Mills
theory play the role of
\textit{moduli}, parameterizing the supersymmetric vacua of the gauge theory.
This has been confirmed in the case of the elliptic
Calogero-Moser system for general Lie algebra
in Refs. \cite{hp3,hp4} and in the case of the elliptic $SU(N)$
RS system for the perturbative limit and some
nonperturbative special point\cite{bm1}.

As for the $C_{n}$ RS system, the spectral curve of can be
generated by Lax matrix $L(\lambda)_{C_{n}}$ as follows

\begin{equation}
\det (L(\lambda )_{C_{n}}-v\cdot Id)=\sum_{j=0}^{2n}\frac{(\sigma (\lambda
))^{^{(j-1)}}\sigma (\lambda +j\gamma )}{(\sigma (\gamma
+\lambda ))^{j}} (-v)^{2n-j}(H_{j})_{C_{n}}=0,
\end{equation}
where
$(\mathcal{H}_{0})_{C_{n}}=(\mathcal{H}_{2n})_{C_{n}}=1$ and
$\ (
\mathcal{H}_{i})_{C_{n}}=$ $(\mathcal{H}_{2n-i})_{C_{n}}$ who Poisson commute

\begin{equation}
\{(\mathcal{H}_{i})_{C_{n}},(\mathcal{H}_{j})_{C_{n}}\}=0,
\mbox{$ \ \ \ \ \ \ \ \
$}i,j=1,...,n.
\end{equation}
This can be deduced by verbose but straightforward
calculation to verify that the
$(\mathcal{H}_{i})_{A_{2n-1}},i=1,...2n$ is the first class
\ Hamiltonian with respect to the constraints Eq.
(\ref{cncon}), using Eq. ( \ref{aninv}), (\ref{cnevo}) and
the first formula of Eq. (\ref{reduce}).

The explicit form of $(\mathcal{H}_{l})_{C_{n}}$ are
\begin{equation}
(\mathcal{H}_{l})_{C_{n}}=\sum_{\stackrel{J\subset
\{1,\ldots ,n\},\,|J|\leq l}{\varepsilon _{j}=\pm 1,\,j\in
J}}\mathrm{\exp }(p_{\varepsilon J})\,F_{\varepsilon
J;\,J^{c}}\,U_{J^{c},\,l-|J|},\;\;\;\;\;\;\;\;l=1,\ldots ,n,
\end{equation}
with
\begin{eqnarray}
p_{\varepsilon J} &=&\sum_{j\in J}\;\varepsilon _{j}\,p_{j},
\nonumber \\ F_{\varepsilon J;\,K}
&=&\,\prod_{\stackrel{j,j^{\prime }\in J}{j<j^{\prime }
}}f^{2}(\varepsilon _{j}x_{j}+\varepsilon _{j^{\prime
}}x_{j^{\prime }})\,\prod_{\stackrel{j\in J}{k\in
K}}f(\varepsilon
_{j}x_{j}+x_{k})f(\varepsilon _{j}x_{j}-x_{k})\prod_{j\in J}f(2\varepsilon
_{j}x_{j}), \\
U_{I,p} &=&\sum_{\stackrel{I^{\prime }\subset I}{|I^{\prime
}|=\lbrack p/2\rbrack }}\prod_{\stackrel{j\in I^{\prime
}}{k\in I\backslash I^{\prime }}
\,}f(x_{jk})f(x_{j}+x_{k})g(x_{jk})g(x_{j}+x_{k})\left\{
\begin{array}{c}
0,\mbox{ \ ($p$ odd)} \\ 1,\mbox{ \ ($p$ even)}
\end{array}
\right..\nonumber
\end{eqnarray}
Here, $\lbrack p/2\rbrack $ denotes the integer part of
$p/2$. As an example, for $C_{2}$ RS model, the independent
Hamiltonian flows $(\mathcal{H}_{1})_{C_{2}}$ and
$(\mathcal{H}_{2})_{C_{2}}$ generated by the Lax matrix $
L_{C_{2}}$ are\cite{kai3}

\begin{eqnarray}
(\mathcal{H}_{1})_{C_{2}}
&=&\mathcal{H}_{C_{2}}=e^{p_{1}}f(2x_{1})
\,f(x_{12})f(x_{1}+x_{2})  \nonumber \\
&&+e^{-p_{1}}g(2x_{1})\,g(x_{12})g(x_{1}+x_{2})	 \nonumber
\\ &&+e^{p_{2}}f(2x_{2})\,f(x_{21})f(x_{2}+x_{1})  \nonumber
\\ &&+e^{-p_{2}}g(2x_{2})\,g(x_{21})g(x_{2}+x_{1}), \\
(\mathcal{H}_{2})_{C_{2}}
&=&e^{p_{1}+p_{2}}f(2x_{1})\,(f(x_{1}+x_{2}))^{2}f(2x_{2})
\nonumber \\
&&+e^{-p_{1}-p_{2}}g(2x_{1})\,(g(x_{1}+x_{2}))^{2}g(2x_{2})
\nonumber \\
&&+e^{p_{1}-p_{2}}f(2x_{1})\,(f(x_{12}))^{2}f(-2x_{2})
\nonumber \\
&&+e^{p_{2}-p_{1}}g(2x_{1})\,(g(x_{12}))^{2}g(-2x_{2})
\nonumber \\
&&+2f(x_{12})\,g(x_{12})\,f(x_{1}+x_{2})g(x_{1}+x_{2}).
\end{eqnarray}

\subsection{Spectral curve of the $BC_{n}$ model}

Similar to the calculation of the $C_n$ case, the spectral curve
of the $BC_{n}$ RS system can be generated by Lax matrix $
L(\lambda)_{BC_{n}} $ as follows

\begin{equation}
\det (L(\lambda )_{BC_{n}}-v\cdot Id)=0.
\end{equation}
The explicit form of the spectral curve is

\[
\det (L(\lambda )_{BC_{n}}-v\cdot Id)=\sum_{j=0}^{2n+1}\frac{(\sigma
(\lambda ))^{^{(j-1)}}\sigma (\lambda +j\gamma )}{(\sigma
(\gamma +\lambda
))^{j}}(-v)^{2n+1-j}(\mathcal{H}_{j})_{BC_{n}}=0,
\]
where
$(\mathcal{H}_{0})_{BC_{n}}=(\mathcal{H}_{2n})_{BC_{n}}=1$
and $\ (
\mathcal{H}_{i})_{BC_{n}}=$ $(\mathcal{H}_{2n+1-i})_{BC_{n}}$ Poisson commute

\begin{equation}
\{(\mathcal{H}_{i})_{BC_{n}},(\mathcal{H}_{j})_{BC_{n}}\}=0,
\mbox{$ \ \ \ \ \
$}\forall i,j\in \{1,...,n\}.
\end{equation}
This can be deduced similarly to the $C_{n}$ case to verify
that $(\mathcal{H}_{i})_{A_{2n}},i=1,...2n$ is the first
class  Hamiltonian with respect to the constraints Eq.
(\ref{bcncon}).

The explicit forms of $(\mathcal{H}_{l})_{BC_{n}}$ are

\begin{equation}
(\mathcal{H}_{l})_{BC_{n}}=\sum_{\stackrel{J\subset
\{1,\ldots ,n\},\,|J|\leq l}{\varepsilon _{j}=\pm 1,\,j\in
J}}\mathrm{\exp } (p_{\varepsilon J})\,F_{\varepsilon
J;\,J^{c}}\,U_{J^{c},\,l-|J|},\;\;\;\;\;\;\;\;l=1,\ldots ,n,
\end{equation}
with

\begin{eqnarray}
p_{\varepsilon J} &=&\sum_{j\in J}\;\varepsilon _{j}\,p_{j},
\nonumber \\ F_{\varepsilon J;\,K}
&=&\,\prod_{\stackrel{j,j^{\prime }\in J}{j<j^{\prime }
}}f^{2}(\varepsilon _{j}x_{j}+\varepsilon _{j^{\prime
}}x_{j^{\prime }})\,\prod_{\stackrel{j\in J}{k\in
K}}f(\varepsilon
_{j}x_{j}+x_{k})f(\varepsilon _{j}x_{j}-x_{k})\prod_{j\in J}f(2\varepsilon
_{j}x_{j})\prod_{j\in J}f(\varepsilon _{j}x_{j}), \\
U_{I,p} &=&\sum_{\stackrel{I^{\prime }\subset I}{|I^{\prime
}|=\lbrack p/2\rbrack }}\prod_{\stackrel{j\in I^{\prime
}}{k\in I\backslash I^{\prime }}
\,}f(x_{jk})f(x_{j}+x_{k})g(x_{jk})g(x_{j}+x_{k})\left\{
\begin{array}{c}
\prod_{i\in I\backslash I^{\prime }}f(x_{i})g(x_{i}),\mbox{ \ ($p$ odd)} \\
\prod_{i^{\prime }\in I^{\prime }}f(x_{i^{\prime }})g(x_{i^{\prime }}),
\mbox{
\ \ ($p$ even)}
\end{array}
\right..\nonumber
\end{eqnarray}

\subsection{Limit to the Calogero-Moser system and Toda system}

The Calogero-Moser system can be achieved by taking
so-called ``the nonrelativistic limit". The procedure is by
rescaling $p_{\mu }\longmapsto
\beta p_{\mu }$, $\gamma \longmapsto \beta \gamma$ and letting $\beta
\longmapsto 0$, followed by making a canonical transformation

\begin{equation}
p_{\mu }\longmapsto p_{\mu }+\gamma \sum_{\Delta \ni \eta
=\mu -\nu }\zeta (\eta \cdot x),
\end{equation}
\noindent here $p_{\mu }=\mu \cdot p,$ such that
\begin{equation}
L\longmapsto Id+\beta L_{CM}+O(\beta ^{2}),
\end{equation}
and

\begin{equation}
\mathcal{H}\longmapsto N+2\beta ^{2}\mathcal{H}_{CM}+O(\beta ^{2}).
\end{equation}
where $N=2n$ for $C_{n}$ model and $N=2n+1$ for $BC_{n}$
model.

$L_{CM}$ can be expressed as

\begin{equation}
L_{CM}=p\cdot H+X,
\end{equation}
where
\begin{equation}
H_{\mu \nu }=\mu \delta _{\mu \nu },  \nonumber \\ X_{\mu
\nu }=\gamma \Phi (x_{\mu \nu },\lambda )(1-\delta _{\mu \nu
}).
\end{equation}
The Hamiltonian $\mathcal{H}_{CM}$ of $\ CM$ model can be
given by

\begin{equation}
\mathcal{H}_{CM}=\frac{1}{2}p^{2}-\frac{\gamma ^{2}}{2}\sum_{\alpha \in
\Delta }\wp(\alpha \cdot x)=\frac{1}{4}trL^{2}+const,\ \ \ \
\label{HCM}
\end{equation}
where $const=-\frac{N(N-1)\gamma ^{2}}{4}\wp(\lambda
).$

All of the above results are identified with the results of
Refs. \cite {op,hp1,bcs2,bcs3,bcs1} up to a suitable choice
of coupling parameters.

Now the degenerate RS spectral curve reduce to

\begin{equation}
\Gamma :\ R(v,\lambda )=~\det (L(\lambda )_{CM}-v\cdot Id)\equiv 0,
\end{equation}
which exactly identified with the spectral curve analyzed in
\cite{hp2,hp3}.

Starting from the CM system to the Toda system is more
directly due to the progress that the limit to Toda for the
general Lie algebra has been studied extensively in
\cite{ino1,hp5,kst}. The main idea is making suitable
scaling limit with the following parameterization
\begin{equation}
\omega _{1}=-i\pi ,\quad \omega _{3}\in \mathbf{R}_{+},\quad \tau \equiv {
\frac{\omega _{3}}{{\omega _{1}}}}=i\omega _{3}/\pi ,  \label{period}
\end{equation}
and shift the dynamical variable $x$
\begin{eqnarray}
x &\rightarrow &Q-2\omega _{3}\,\delta \,\rho ^{\vee
},~~~~~~~p\rightarrow P,
\nonumber \\
\lambda &\rightarrow &\log Z-\omega _{3},\quad Z\in \mathbf{R}_{+},
\label{shift}
\end{eqnarray}
in which $h_{\mathcal{G}}$ is the Coxeter number for the
corresponding root system $\mathcal{G}$, $\rho ^{\vee }$ the
dual of the Weyl vector defined as $\rho ^{\vee
}={\frac{1}{2}}\sum_{\alpha \in \Delta _{+}}2\alpha /\alpha
^{2} $ and $\delta $ satisfies $\delta \leq 1/h_{\mathcal{G}}$.

For convenience, we give the basics of these root system as
showed in Table \ref{table:1}.
\begin{table}[h]
\begin{center}
\begin{tabular}{|c|c|c|c|c|c|}
\hline
$\mathcal{G}$ & $\mbox{all roots}$ & $\mbox{simple roots}$
$\Pi $ & $h_{
\mathcal{G}}$ & $\mbox{dual
Weyl vector}$ $\rho ^{\vee }$ & $\mbox{vector weights}$ \\
\hline $A_{n-1}$ & $
\begin{array}{c}
\pm e_{i}\pm e_{j}, \\
1\leq i,j\leq n, \\ i\neq j
\end{array}
$ & $
\begin{array}{c}
e_{i}-e_{i+1}, \\ i=1,...,n-1
\end{array}
$ & $n$ & $\sum_{j=1}^{n}(n-j)e_{j}$ & $
\begin{array}{c}
e_{i}, \\ i=1,...,n
\end{array}
$ \\ \hline $C_{n}$ & $
\begin{array}{c}
\pm e_{i}\pm e_{j},\pm 2e_{i}, \\
1\leq i,j\leq n, \\ i\neq j
\end{array}
$ & $
\begin{array}{c}
e_{i}-e_{i+1},2e_{n}, \\ i=1,...,n-1
\end{array}
$ & $2n$ & $\sum_{j=1}^{n}(n+\frac{1}{2}-j)e_{j}$ & $
\begin{array}{c}
e_{i},-e_{i}, \\ i=1,...,n
\end{array}
$ \\ \hline $BC_{n}$ & $
\begin{array}{c}
\pm e_{i}\pm e_{j},\pm 2e_{i},\pm e_{i} \\
1\leq i,j\leq n, \\ i\neq j
\end{array}
$ & $
\begin{array}{c}
e_{i}-e_{i+1},e_{n}, \\ i=1,...,n-1
\end{array}
$ & $2n+1$ & $
\begin{array}{c}
\sum_{j=1}^{n}(n+1-j)e_{j} \\
(define)
\end{array}
$ & $
\begin{array}{c}
e_{i},-e_{i},0, \\ i=1,...,n
\end{array}
$ \\ \hline
\end{tabular}
\caption{Root system of $A_{n-1}$, $C_{n}$ and $BC_{n}$ types}
\label{table:1}
\end{center}
\end{table}

As for the $C_{n}$ model, selecting $\rho^{\vee }=
\rho _{C_{n}}^{\vee }$, $\gamma =im\,e^{\omega _{3}\delta }$, one have the
non-affine $C_{n}$ Toda model from the Hamiltonian of the CM
model Eq. (\ref{HCM})
\begin{equation}
\mathcal{H}_{C_{n}}^{\mathit{Toda}}={\frac{1}{2}}P^{2}+m^{2}
\sum_{j=1}^{n-1}e^{Q_{j}-Q_{j+1}}+m^{2}\,e^{2Q_{n}},
\end{equation}
for \ $\delta <1/h_{C_{n}}$ and $C_{n}^{(1)}$ Toda model
\begin{equation}
\mathcal{H}_{C_{n}^{(1)}}^{\mathit{Toda}}={\frac{1}{2}}
P^{2}+m^{2}e^{-2Q_{1}}+m^{2}\sum_{j=1}^{n-1}e^{Q_{j}-Q_{j+1}}+m^{2}e^{2Q_{n}},
\end{equation}
for \ $\delta =1/h_{C_{n}}$.

The same holds for the $BC_{n}$ model. Selecting $\rho^{\vee }=
\rho _{BC_{n}}^{\vee }$, $\gamma =im\,e^{\omega _{3}\delta }$, one have the
non-affine $B_{n}$ Toda model from the Hamiltonian of the CM
model Eq. (\ref{HCM})
\begin{equation}
\mathcal{H}_{B_{n}}^{\mathit{Toda}}={\frac{1}{2}}P^{2}+m^{2}
\sum_{j=1}^{n-1}e^{Q_{j}-Q_{j+1}}+m^{2}\,e^{Q_{n}},
\end{equation}
for \ $\delta <1/h_{BC_{n}}$ and $BC_{n}$ Toda model
\begin{equation}
\mathcal{H}_{BC_{n}}^{\mathit{Toda}}={\frac{1}{2}}
P^{2}+m^{2}e^{-2Q_{1}}+m^{2}\sum_{j=1}^{n-1}e^{Q_{j}-Q_{j+1}}+m^{2}e^{Q_{n}},
\end{equation}
for \ $\delta =1/h_{BC_{n}}$.

If we use the following gauge for $\Phi (x,\lambda )$
\cite{Kri}
\begin{equation}
\Phi (x,\lambda )\rightarrow {\frac{\sigma (x+\lambda )}{{\sigma (\lambda
)\sigma (x)}}}\exp ({\zeta (\lambda )x}),  \label{funx}
\end{equation}
which does not destroy the validity\cite{kst} for the Lax
pair, We have the following limit for $\gamma \Phi (\alpha
\cdot x,\lambda )$

\begin{equation}
\begin{array}{llll}
\gamma \Phi (\alpha \cdot x,\lambda ) & \rightarrow & -m\exp ({\frac{{\alpha
\cdot Q}}{{2}}}) & \mbox{for}\quad \alpha \in \Pi \quad (\delta \leq 1/h_{
\mathcal{G}}), \\
& \rightarrow & {mZ}\exp (-{\frac{{\alpha \cdot Q}}{{2}}}) &
\mbox{for}\ \ \
\ \alpha =\alpha _{h}\quad (\delta =1/h_{\mathcal{G}}), \\
& \rightarrow & 0 & \mbox{otherwise}, \\
\gamma \Phi (-\alpha \cdot x,\lambda ) & \rightarrow & m\exp ({\frac{{\alpha
\cdot Q}}{{2}}}) & \mbox{for}\quad \alpha \in \Pi \quad (\delta \leq 1/h_{
\mathcal{G}}), \\
& \rightarrow & -{\frac{m}{Z}}\exp (-{\frac{{\alpha \cdot
Q}}{{2}}}), &
\mbox{for}\ \ \ \ \alpha =\alpha _{h}\quad (\delta =1/h_{\mathcal{G}}), \\
& \rightarrow & 0 & \mbox{otherwise}.
\end{array}
\end{equation}
So the Lax operator now reads
\begin{equation}
L_{\mathit{Toda}}=P\cdot H-im\sum_{\alpha \in \Pi }\exp
({\frac{{\alpha
\cdot Q}}{2}})\lbrack E(\alpha )-E(-\alpha )\rbrack +im\exp ({\frac{{\alpha
_{0}\cdot Q}}{2}})\lbrack ZE(-\alpha _{0})-Z^{-1}E(\alpha _{0})\rbrack ,
\label{todalax}
\end{equation}
where $E(\alpha )_{\mu \nu }=\delta _{\mu -\nu ,\alpha }$.
This Lax operator holds for all the root system of
$A_{n-1}(A_{n-1}^{(1)}),$ $C_{n}(C_{n}^{(1)}),$ $B_n(BC_{n})$ and coincide with the standard
form given in \cite{op}. It is not difficult to find that
the parameter $Z$ now plays the role of a spectral parameter
for the affine Toda model based on ${\mathcal{G}}^{(1)}$.
When we refer to the Toda models based on a finite Lie
algebra $\mathcal{G}$, we should only drop the terms
containing the affine root $\alpha _{0}$.

So the degenerate spectral curve for the  Toda $A_{n-1}^{(1)}$
, $C_{n}^{(1)}$ and  $BC_{n}(A_{2n}^{(2)})$ systems can be
defined

\begin{equation}
\Gamma \ :\ R(v,\lambda )=~\det (L(\lambda )_{\mathit{Toda}}-v\cdot
Id)\equiv 0,
\end{equation}
which is identical to the one given in \cite{Martin,Taka}.

\section{Degenerate Cases}

\setcounter{equation}{0} \label{degen} Let us now consider the other various
special degenerate cases. As is well known, if one or both
the periods of Weierstrass sigma function $\sigma (x)$
become infinite, there will occur three degenerate cases
associated with trigonometric, hyperbolic and rational
systems. The degenerate limits of the functions $\Phi
(x,\lambda )$, $
\sigma (x)$ and $\zeta (x)$ will give corresponding Lax pairs which include
spectral parameter. Moreover, when the \ spectral parameter
value on certain limit, the Lax pairs without spectral
parameter will be derived.

\subsection{Trigonometric limit}

The limit can be obtained by sending \ $\omega _{3}$ \ to
$i\infty $ with $
\omega _{1}=\frac{\pi }{2},$ so that

\begin{eqnarray}
\sigma (x) &\rightarrow &e^{\frac{1}{6}x^{2}}\sin x,  \nonumber \\
\zeta (x) &\rightarrow &\cot x+\frac{1}{3}x,  \label{Tlimit}
\end{eqnarray}
and the function $\Phi (x,\lambda )\equiv \frac{\sigma
(x+\lambda )}{\sigma (x)\sigma (\lambda )}$ reduce to

\begin{equation}
\Phi (x,\lambda )\rightarrow (\cot \lambda -\cot x)e^{\frac{1}{3}xu}.
\end{equation}
By replacing the corresponding \ functions $\Phi (x,\lambda
)$\ ,\ $\ \sigma (x)$ and $\zeta (x)$ to the form given
above for the Lax pairs$,$ we will get the corresponding
spectral parameter dependent Lax pairs. For the simplicity,
we notice that the exponential part of the above functions
can be removed by applying \ suitable ''gauge''
transformation of the Lax matrix on which condition the
functions can be valued as follows:

\begin{eqnarray}
\sigma (x) &\rightarrow &\sin x,  \nonumber \\
\zeta (x) &\rightarrow &\cot x, \\
\Phi (x,\lambda ) &\rightarrow &(\cot \lambda -\cot x).	 \nonumber
\end{eqnarray}

As for the spectral parameter independent Lax pair,
furthermore, we can take the limit $\lambda \rightarrow
i\infty $, so the function

\begin{equation}
\Phi (x,\lambda )\rightarrow \frac{1}{\sin x},
\end{equation}
while the corresponding Lax matrix become to
\begin{equation}
L_{\mu \nu }=e^{\nu \cdot p}b_{\nu }\frac{\sin \gamma }{\sin
((\mu -\nu )\cdot x+\gamma )},	\label{trigL}
\end{equation}
which are exactly the same as the spectral parameter
independent Lax matrix given in \cite{Che00}.

\subsection{Hyperbolic limit}

In this case, the periods can be chosen as by sending by
sending \ $\omega
_{1}$ \ to $i\infty $ with $\omega _{3}=\frac{\pi }{2},$ so following all
the procedure in achieving the result of trigonometric case,
we can \ find the \ hyperbolic \ Lax pairs by simple
replacement of the functions appeared in trigonometric Lax
pair \ as follows:
\begin{eqnarray}
\sin x &\rightarrow &\sinh x,  \nonumber \\
\cos x &\rightarrow &\cosh x, \\
\cot x &\rightarrow &\coth x.  \nonumber
\end{eqnarray}
The same as for the trigonometric case, we can get the Lax
pairs with and without spectral parameter.

\subsection{Rational limit}

As far as the form of the Lax pair for the rational-type
system is concerned, we can achieve it by making the
following substitutions
\begin{eqnarray}
\sigma (x) &\rightarrow &x,  \nonumber \\
\zeta (x) &\rightarrow &\frac{1}{x}, \\
\Phi (x,\lambda ) &\rightarrow &\frac{1}{x}+\frac{1}{\lambda }.	 \nonumber
\end{eqnarray}
for the spectral parameter dependent \ Lax pair, while
furthermore, taking the limit $\lambda \rightarrow i\infty
$, we can obtain the spectral parameter independent \ Lax
pair. The explicit form of Lax matrix without spectral
parameter is
\begin{equation}
L_{\mu \nu }=e^{\nu \cdot p}b_{\nu }\frac{\gamma }{(\mu -\nu
)\cdot x+\gamma }.  \label{ratiL}
\end{equation}
which completely coincide with the spectral parameter
independent \ Lax matrix given in \cite{Che00}.

\vspace{0.2truecm}

\textbf{Remark:} As for the various degenerate cases for the CM and Toda
systems, one can follow the same procedure as for the RS
model(please refer to Eq. (\ref{Tlimit})-(\ref{ratiL})).

\section{Concluding Remarks}

\setcounter{equation}{0} \label{summ} In this paper, we have proposed the
Lax pairs for elliptic $C_{n}$ and $BC_{n}$ RS models. The
spectral parameter dependent and independent Lax pairs for
the trigonometric, hyperbolic and rational systems can be
derived as the degenerate limits of the elliptic potential
case. The spectral curves of these systems are given and
shown by depicted by the complete sets of involutive
constant integrals of motion.
They would  be related to the
integrable system associated with  the 5-dimensional gauge
theory\cite{n,bm1}. In the
nonrelativistic limit(scaling limit), the system leads to
CM(Toda) systems associated with the root systems of $C_{n}$
and $BC_{n}$. There are still
many open problems, for example, it seems to be a
challenging subject to carry out the Lax pairs with as many
independent coupling constants as independent Weyl orbits in
the set of roots, as done for the Calogero-Moser systems(see
Refs. \cite{op}, \cite{bcs}
-\cite{hm}). What is also interesting may generalize the results obtained
in this paper to the systems associated with all of other
Lie Algebras even to those associated with all the finite
reflection groups\cite{bcs1}. Moreover, The issue for
getting the r-matrix structure for these system is deserved
due to the success of calculating for the \ trigonometric
$BC_{n}$ RS system by Avan et al. in \cite{Avan}.

\section*{Acknowledgement}

We would like to thank professors K. J. Shi and L. Zhao for
useful and stimulating \ discussions. This work has been
supported financially by the National Natural Science
Foundation of China. W.-L. Yang has been supported by the
Alexander von Houmblodt foundation.

\section*{Appendix}

\renewcommand{\theequation}{A.\arabic{equation}}
\setcounter{equation}{0} In
this appendix we prove the identity Eq. (\ref{laxinverse})
and then derive the relation between the Lax operator
$L(\lambda )$ and its inverse of $L(\lambda )^{-1}$.

Using the result given in \ \cite{r1} of Eq. (B5), we have
the following conclusion:

\noindent \textit{Let}
\begin{equation}
C_{ij}=\frac{\;\sigma (q_{i}-r_{j}+\lambda )}{\;\sigma
(q_{i}-r_{j}+\mu )},
\mbox{$ \ \ \ \ \ \ \ \ \ \ $}i,j=1,...,N,
\end{equation}
\textit{then one has }

\begin{eqnarray}
\det (C) &=&\sigma (\lambda -\mu )^{N-1}\sigma (\lambda +(N-1)\mu +\Sigma )
\nonumber \\
&&\times \prod_{i<j}\sigma (q_{i}-q_{j})\sigma
(r_{j}-r_{i})\prod_{i,j}\frac{ 1}{\sigma (q_{i}-r_{j}+\mu
)},  \label{A1}
\end{eqnarray}
\textit{where}
\begin{equation}
\Sigma =\sum_{i=1}^{N}(q_{i}-r_{j}).
\end{equation}

\vspace{0.2cm} So it is straightforward to compute the inverse of matrix $C$
,
\begin{eqnarray}
(C^{-1})_{ij} &=&\mbox{the cofactor of }C\mbox{ with respect
to }C_{ji}
\nonumber \\
&=&\frac{\sigma (\lambda +(N-2)\mu +q_{i}-q_{j})}{\sigma
(\lambda -\mu )\sigma (\lambda +(N-1)\mu )\sigma
(q_{i}-q_{j}-\mu )}  \nonumber \\ &&\times
\frac{\prod_{l}\sigma (q_{j}-q_{l}+\mu )\prod_{l}\sigma
(q_{i}-q_{l}-\mu )}{\prod_{k\neq i}\sigma
(q_{i}-q_{k})\prod_{k\neq j}\sigma (q_{j}-q_{k})}.
\label{inc}
\end{eqnarray}
From Eq. (\ref{laxA}), we have
\begin{eqnarray}
L(\lambda ) &=&\sum_{i,j=1}^{N}\frac{\Phi
(x_{ij}+\gamma ,\lambda )}{
\Phi (\gamma ,\lambda )}\mathrm{\exp }(p_{j})b_{j}E_{ij}  \nonumber \\
&=&\frac{1}{\Phi (\gamma ,\lambda
)}\sum_{i,j=1}^{N}\frac{\sigma (x_{ij}+\gamma +\lambda
)}{\sigma (x_{ij}+\gamma )\sigma (\lambda )}\mathrm{
\exp }(p_{j})b_{j}  \nonumber \\
&=&\frac{1}{\Phi (\gamma ,\lambda
)}\sum_{i,j=1}^{N}G_{ij}\mathrm{\exp } (p_{j})b_{j},
\end{eqnarray}
where
\[
G_{ij}:=\Phi (x_{ij}+\gamma ,\lambda )=\frac{\sigma
(x_{ij}+\gamma +\lambda )}{\sigma (x_{ij}+\gamma )\sigma
(\lambda )},
\]
with the help of Eq. (\ref{inc}), one has
\begin{eqnarray}
(G^{-1})_{ij} &=&\frac{\sigma (\lambda +(N-1)\gamma
+x_{ij})}{\sigma (\lambda +N\gamma )\sigma (x_{ij}-\gamma )}
\nonumber \\ &&\times \frac{\prod_{k}\sigma (x_{jk}+\gamma
)\prod_{k}\sigma (x_{ik}-\gamma )}{\prod_{k\neq i}\sigma
(x_{ik})\prod_{k\neq j}\sigma (x_{jk})},
\end{eqnarray}
So that
\begin{eqnarray}
L(\lambda )^{-1}{}_{ij} &=&\Phi (\gamma ,\lambda
)(G^{-1})_{ij}\ b_{j}^{-1}
\mathrm{\exp }(-p_{i}))E_{ij}  \nonumber \\
&=&\frac{-\sigma (\gamma )^{2}\,\sigma (\lambda +\gamma
)\sigma (\lambda +(N-1)\gamma +x_{ij})}{\sigma (\lambda
)\sigma (\gamma )\sigma (\lambda +N\gamma )\sigma
(x_{ij}-\gamma )}  \nonumber \\ &&\times \mathrm{\exp
}(-p_{i})\prod_{k\neq j}\frac{\sigma (x_{jk}+\gamma )}{
\sigma (x_{jk})}  \nonumber \\
&=&\frac{\sigma (\gamma +\lambda )\,\sigma (\lambda
+(N-1)\gamma )}{\sigma (\lambda )\sigma (\lambda +N\gamma )}
\nonumber \\ &&\times \frac{\;\Phi (x_{ij}-\gamma
,\lambda +N\gamma )}{\Phi (-\gamma ,\lambda +N\gamma
)}\mathrm{\exp }(-p_{i})b_{j}^{^{\prime }}E_{ij}.
\end{eqnarray}
By comparing the forms of $L(\lambda )$ and $L(\lambda
)^{-1}{}_{ij}$, we find \ $L(\lambda )^{-1}{}_{ij}$ can be
expressed with $L(\lambda )$ as:
\begin{equation}
L(\lambda )^{-1}{}_{ij}=L(\lambda )_{ij}\left|{}_{\gamma
\rightarrow -\gamma ,\lambda \rightarrow \lambda
+N\gamma}\right. \times \frac{\sigma (\gamma +\lambda
)\,\sigma (\lambda +(N-1)\gamma )}{\sigma (\lambda )\sigma
(\lambda +N\gamma )}\mathrm{\exp }(-p_{i}-p_{j}).
\label{Inll}
\end{equation}


\vspace{1pt}

\begin{thebibliography}{99}
\bibitem{Che00}	 K. Chen, B.Y. Hou, W.-L. Yang, Integrability of the $C_{n}$
and $BC_{n}$ Ruijsenaars-Schneider models,
\texttt{hep-th/0006004}, J. Math. Phys. \textbf{41} (2000) 8132.

\bibitem{h1}  K. Hasegawa, Ruijsenaars' commuting difference operators as
commuting transfer matrices, Commun. Math. Phys.
\textbf{187} (1997) 289.

\bibitem{nksr}	F.W. Nijhoff, V.B. Kuznetsov, E.K. Sklyanin and O. Ragnisco,
Dynamical $r$-matrix for the elliptic Ruijsenaars-Schneider
system, J. Phys. A: Math. Gen. \textbf{29}, L333 (1996).

\bibitem{gm}  A. Gorsky, A. Marshakov, Towards effective topological gauge
theories on spectral curves, Phys. Lett. \textbf{B375} (1996) 127.

\bibitem{n}  N. Nekrasov, Five-dimension gauge theories and relativistic
integrable systems, \texttt{hep-th/9609219}, Nucl. Phys.
\textbf{B531} (1998) 323.

\bibitem{bm1}  H.W. Braden, A. Marshakov, A. Mironov, and A. Morozov, The
Ruijsenaars-Schneider model in the context of Seiberg-Witten
theory, \texttt{ hep-th/9902205}, Nucl. Phys. \textbf{B558} (1999) 371.

\bibitem{r1}  S.N.M. Ruijsenaars, Complete integrability of relativistic
Calogero-Moser systems and elliptic function identities,
Commun. Math. Phys. \textbf{110} (1987) 191.


\bibitem{r2}  S.N.M. Ruijsenaars, Action-angle maps and scattering theory
for some finite-dimensional integrable systems, Commun.
Math. Phys. \textbf{ 115} (1988) 127.

\bibitem{op}  M.A. Olshanetsky and A.M. Perelomov, Classical integrable
finite-dimensional systems related to Lie algebras, Phys.
Rep. \textbf{71}, (1981) 314; A.M. Perelomov,
\textit{Integrable Systems of Classical Mechanics
and Lie Algebras} (Birkh\"{a}user, Boston, 1990).

\bibitem{in}  VI. Inozemtsev, Lax representation with spectral parameter on
a torus for integrable particle systems, Lett. Math. Phys.
\textbf{17} (1989) 11,

\bibitem{hp1}  E. D'Hoker and D.H. Phong, Calogero-Moser Lax pairs with
spectral parameter for general Lie algebras,
\texttt{hep-th/9804124, } Nucl. Phys. \textbf{B530} (1998)
537.

\bibitem{bcs}  A.J. Bordner, E. Corrigan and R. Sasaki, Calogero-Moser
Models I: A new formulation, \texttt{hep-th/9805106}, Prog.
Theor. Phys. \textbf{100} (1998) 1107.

\bibitem{bcs2}	A.J. Bordner, R. Sasaki and K. Takasaki, Calogero-Moser
Models II: Symmetries and foldings, \texttt{hep-th/9809068,
} Prog. Theor. Phys. \textbf{101} (1999) 487.

\bibitem{bcs3}	A.J. Bordner and R. Sasaki, Calogero-Moser Models III:
Elliptic potentials and twisting, \texttt{hep-th/9812232,} Prog. Theor.
Phys. \textbf{101} (1999) 799.

\bibitem{bcs1}	A.J. Bordner, E. Corrigan and R. Sasaki,
Generalized Calogero-Moser models and universal Lax pair operators,
\texttt{ hep-th/9905011}, Prog. Theor. Phys. \textbf{102}
(1999) 499.

\bibitem{hm}  J.C. Hurtubise and E. Markman, Calogero-Moser systems and
Hitchin systems, \texttt{math/9912161}.


\bibitem{bms}  A.J. Bordner, N.S. Manton and R. Sasaki,
Calogero-Moser models V: Supersymmetry and quantum Lax pair,
\texttt{ hep-th/9910033,} Prog. Theor. Phys. \textbf{103}
(2000) 463.


\bibitem{kps} S.P. Khastgir, A.J. Pocklington and R.
Sasaki, Quantum Calogero-Moser models: Integrability for all
root systems, \texttt{hep-th/0005277}, J. Phys. \textbf{A33}
(2000) 9033.


\bibitem{op1}  M.A. Olshanetsky and A.M. Perelomov, Quantum
integrable  systems related to Lie algebras, Phys.
Rep. \textbf{94}, (1983) 313.

\bibitem{aru}  G.E. Arutyunov, S.A. Frolov and P.B. Medvedev, Elliptic
Ruijsenaars-Schneider model from the cotangent bundle over
the two-dimensional current group, \texttt{hep-th/9608013,}
J. Math. Phys. \textbf{38 }(1997) 5682.

\bibitem{ko1}  Y. Komori, K. Hikami, Conserved operators of the generalized
elliptic Ruijsenaars models, J. Math. Phys. \textbf{39}
(1998) 6175.

\bibitem{ko2}  Y. Komori, Theta functions associated with the affine root
systems and the elliptic Ruijsenaars operators, \texttt{math.QA/9910003}.

\bibitem{di}  J.F. van Diejen, Integrability of difference Calogero-Moser
systems, J. Math. Phys. \textbf{35} (1994) 2983.

\bibitem{di1}  J.F. van Diejen, Commuting difference operators with
polynomial eigenfunctions, Compositio. Math. \textbf{95}
(1995) 183.

\bibitem{h2}  K. Hasegawa, T. Ikeda, T. Kikuchi, Commuting difference
operators arising from the elliptic $C_{2}^{(1)}$-face
model, J. Math. Phys. \textbf{40} (1999) 4549.


\bibitem{bc}  M. Bruschi and F. Calogero, The Lax pair representation for an
integrable class of relativistic dynamical systems, Commun.
Math. Phys. \textbf{109} (1987) 481.

\bibitem{kz}  I. Krichever and A. Zabrodin, Spin generalization of the
Ruijsenaars-Schneider model, the non-Abelian $2D$ Toda
chain, and representations of the Sklyanin algebra, Usp.
Math. Nauk, \textbf{50:6} (1995) 3.

\bibitem{s1}  Y.B. Suris, Why are the rational and hyperbolic
Ruijsenaars-Schneider hierarchies governed by the same
$R$-operators as the Calogero-Moser ones?,
\texttt{hep-th/9602160}.

\bibitem{s2}  Y.B. Suris, Elliptic Ruijsenaars-Schneider and Calogero-Moser
hierarchies are governed by the same $r$-matrix,
\texttt{solv-int/9603011, } Phys. Lett. \textbf{A225}
(1997) 253.

\bibitem{kai3}	K. Chen, B.Y. Hou and W.-L. Yang, The Lax pair for $C_{2}$
-type Ruijsenaars-Schneider model, \texttt{hep-th/0004006}, Chin. Phys.
(in press).

\bibitem{Avan}	J. Avan and G. Rollet, $BC_{n}$ Ruijsenaars-Schneider
models: R-matrix structure and Hamiltonians,
\texttt{hep-th/0008174}.


\bibitem{ohta}	Y. Ohta, Instanton correction of prepotential in Ruijsenaars
model associated with N=2 SU(2) Seiberg-Witten Theory,
\texttt{hep-th/9909196 },  J. Math. Phys. \textbf{41} (2000)
4541.

\bibitem{Avan1}	 J. Avan, Classical dynamical $r$-matrices for
Calogero-Moser systems and their generalizations, \texttt{q-alg/9706024},
PAR-LPTHE 97-24.

\bibitem{kns}  V.B. Kuznetsov, F.W. Nijhoff and E.K. Sklyanin, Separation of
variables for the Ruijsenaars system, Commun. Math. Phys.
\textbf{189} (1997) 855.

\bibitem{r3}  S.N.M. Ruijsenaars and H. Schneider, A new class of integrable
systems and its relation to solitons, Ann. Phys.
\textbf{170} (1986) 370.

\bibitem{Dirac}	 Paul A.M. Dirac, \textit{Lectures on Quantum Physics}
(Yeshiva University, New York, 1964).

\bibitem{mir}  A. Mironov and Morozov, Double elliptic systems: problems and
perspectives, \texttt{hep-th/0001168}.

\bibitem{mir1} A. Mironov, Seiberg-Witten theory and duality in integrable
systems, \texttt{hep-th/0011093}

\bibitem{hp2}  E. D'Hoker and D.H. Phong, Lectures on supersymmetric
Yang-Mills theory and integrable Systems,
\texttt{hep-th/9912271}, UCLA/99/TEP/28.

\bibitem{Marbook}  A. Marshakov, \textit{Seiberg-Witten Theory and
Integrable Systems} (World Scientific, Singapore, 1998).

\bibitem{hp3}  E. D'Hoker and D.H. Phong, Spectral curves for
Super-Yang-Mills with adjoint hypermultiplet for general Lie
algebras,
\texttt{hep-th/9804126,} Nucl. Phys. \textbf{B534} (1998) 697.

\bibitem{hp4}  E. D'Hoker and D.H. Phong, Lax pairs and spectral curves for
Calogero-Moser and spin Calogero-Moser systems,
\texttt{hep-th/9903002,} UCLA/99/TEP/3, Columbia/99/Math.

\bibitem{ino1}	V. I. Inozemtsev, The finite Toda lattices, Comm. Math.
Phys. \textbf{121} (1989) 628.

\bibitem{hp5}  E. D'Hoker and D.H. Phong, Calogero-Moser and Toda systems
for twisted and untwisted affine Lie Algebras,
\texttt{hep-th/9804125, } Nucl. Phys. \textbf{B530 }(1998)
611.

\bibitem{kst}  S.P. Khastgir, R. Sasaki and K. Takasaki, Calogero-Moser
Models IV: Limits to Toda theory, \texttt{hep-th/9907102},
Prog. Theor. Phys. \textbf{102} (1999) 749.

\bibitem{Kri}  I.M. Krichever, Elliptic solutions of
the Kadomtsev-Petviashvili equation and integrable systems
of particles, Funct. Anal. Appl. \textbf{14} (1980) 282.

\bibitem{Martin}  E. Martinec and N. Warner,
Integrable Systems and supersymmetric gauge theory,
\texttt{hep-th/9509161, } Nucl. Phys. \textbf{B459} (1996)
97.

\bibitem{Taka}	K. Takasaki, Whitham deformation of Seiberg-Witten curves
for classical gauge groups, \texttt{hep-th/9901120}, Int. J. Mod. Phys.
\textbf{A15} (2000) 3635.

\end{thebibliography}
\end{document}